\documentclass[11pt]{article}
\usepackage{amsmath,amssymb,amsthm,cite}
\title{Singularity Confinement and Algebraic Integrability}
\author{\\
\large{S. \textsc{Lafortune}}\footnote{lafortus@math.arizona.edu}
     \large{ and A. \textsc{Goriely}}\footnote{goriely@math.arizona.edu} \\
\small
\begin{tabular}{c}
     Department of Mathematics\\
     University of Arizona,\\
     Tucson, AZ, 85721--0089, USA\\
\end{tabular}\\
}
\numberwithin{equation}{section}

\begin{document}
\maketitle
\def\D{{\mbox{D}}}
\def\Lc{\overline{L}_c}
\def\Lcv{\overline{L}_{\c}}
\def\Lcs{\overline{L}_{c^*}}
\def\Lcsv{\overline{L}_{\c^*}}
\def\f{{\bf{f}}}
\def\s{{\bf{s}}}
\def\c{{\bf{c}}}
\def\Ima{\rm Im}
\def\Re{\rm Re}
\def\xb{{\bf{x}}}
\def\xd{x_{n-1}}
\def\xu{x_{n+1}}
\def\xn{x_n}
\def\yd{y_{n-1}}
\def\yu{y_{n+1}}
\def\yn{y_n}
\def\xbd{{\bf{x}}_{n-1}}
\def\xbu{{\bf{x}}_{n+1}}
\def\xbn{{\bf{x}}_n}
\def\ap{a_{1,n}}
\def\apu{a_{1,n+1}}
\def\apd{a_{1,n-1}}
\def\as{a_{2,n}}
\def\asu{a_{2,n+1}}
\def\asd{a_{2,n-1}}
\def\u{u_n}
\def\v{v_n}
\def\a{a_n}
\def\b{b_n}
\def\d{d_n}
\def\vu{v_{n+1}}
\def\vd{v_{n-1}}
\def\au{a_{n+1}}
\def\ad{a_{n-1}}
\def\cu{c_{n+1}}
\def\cd{c_{n-1}}
\def\bu{b_{n+1}}
\def\bd{b_{n-1}}
\def\du{d_{n+1}}
\def\dd{d_{n-1}}
\def\uu{u_{n+1}}
\def\ud{u_{n-1}}
\def\SI{S_I({\bf{f}})}
\def\SII{S_{I\hspace{-.05cm}I}({\bf{f}})}
\def\SIIw{S_{I\hspace{-.05cm}I}}
\newcommand{\ts}{\textstyle}
\newcommand{\ds}{\displaystyle}
\newcommand{\dt}{\partial_{t}}
\newcommand{\duk}{\partial_{u_{k}}}
\newcommand{\dvk}{\partial_{v_{k}}}
\newcommand{\dun}{\partial_{u_{n}}}
\newcommand{\dvn}{\partial_{v_{n}}}
\newcommand{\vect}[2]{\left( \begin{array}{c} #1 \\#2 \end{array} \right)}
\newtheorem{theorem}{Theorem}

\newcommand{\mat}[4]{\ensuremath \left(  \begin{array}{cc} #1 & #2 \\
#3 & #4 \end{array} \right) } \vspace{-1cm}

\abstract{ \hrule\vskip 12pt Two important notions of
integrability for discrete mappings are algebraic integrability
and singularity confinement, have been used for discrete mappings.
Algebraic integrability is related to the existence of
sufficiently many conserved quantities whereas singularity
confinement is associated with the local analysis of
singularities.  In this paper, the relationship between these two
notions is explored for birational autonomous mappings. Two types
of results are obtained: first, algebraically integrable mappings
are shown to have the singularity confinement property. Second, a
proof of the non-existence of algebraic conserved quantities of
discrete systems based on the lack of confinement property is
given.
\\}
\markboth{}{Singularity Confinement and Algebraic Integrability}
\hrule
\nopagebreak

\section{INTRODUCTION}\label{s1}

One of the first notion of integrability was introduced by Liouville in the
nineteenth century for Hamiltonian systems in classical mechanics. This
   type of integrability is based on quantities whose
value do not change in time, the so-called {\it{constants of the
motion}} or {\it{first integrals}}. For instance, in many mechanical systems
the total energy and the linear and angular momenta  are
conserved through the dynamics and are examples of such constants.
Liouville's fundamental contribution was to prove that if a
given Hamiltonian systems admits enough first integrals, it can be solved
explicitly by quadratures. More precisely, a
$2n$-dimensional Hamiltonian system is said to be \textit{Liouville-integrable}
if it admits $n$ functionally independent first integrals in
involution (that is
their Poisson Brackets commute) and such systems can  be integrated by
quadrature and enjoy a particularly simple topology (the flow lives
on products of tori and cylinders).
   For systems of $n$ first-order ordinary differential equations (ODEs)
\textit{algebraic integrability} is defined  as the existence
$(n-1)$ functionally  independent first integrals that are
algebraic functions of the dependent variables. The existence of
such first integrals is important for  integrability as they can
be used  to reduce the dimensionality of the system.

In the particular case of Hamiltonian systems, algebraic
integrability coincides with the notion of superintegrability.
Indeed, for a $2n$-dimensional Hamiltonian system, the $2n-1$
first integrals needed for the property of algebraic integrability
are much more than the $n$ ones needed for complete
Liouville-Arnold integrability.

A  different notion of integrability was introduced by Painlev\'e
in the beginning of the 20th century \cite{1,2}. Although the aim
of Painlev\'e was not to define a notion of integrability but
rather to build new functions, his property is today widely used
for the detection of integrable systems. An ODE is said to possess
the {\it{Painlev\'e property}} \cite{3,4,Goriely02} if its general
solution is single-valued in its maximal domain of analytic
continuation. The restrictions the Painlev\'e property impose on
the solutions are so strong that an ODE exhibiting it may be
considered for all practical purpose integrable. Many formal links
between the Painlev\'e property and other notions of integrability
have been established (see for example \cite{Goriely02,6}).
Despite the fact that there is no general algorithmic way to
obtain sufficient conditions for a given ODE to have the
Painlev\'e property,  necessary conditions can be derived. The
algorithmic procedure to obtain such conditions is  known as the
{\it{Painlev\'e test} or \textit{singularity analysis}}  and is
based on a local analysis of the solutions around  movable
isolated singularities.


The notions of algebraic integrability and singularity analysis of
the solutions can be extended to finite difference equations. One
of the most effective way to perform singularity analysis in this
context is given by {\it{singularity confinement}}
\cite{6,7,7b,7a,8,9,10,11,12,12a,13} which, like the Painlev\'e
test for ODEs, imposes conditions on the singularities of the
solutions. Although there exist many striking particular examples
which indicate that  singularity confinement  is
closely related to other notions of integrability, there is no clear-cut result
which establishes this relationship in a formal and general
framework. This paper explores the relationship between the notion of algebraic
integrability and the property of singularity confinement in discrete
mappings.

In the case of systems of ODEs, Yoshida \cite{yo83,yo83b,Goriely02}
proved that the degree of a rational first integral is closely related to some
exponents that can be obtained through the application of the
Painlev\'e test. In
this paper, we establish  an equivalent result for discrete systems by  showing
that  singularity confinement analysis provides necessary conditions for the
existence of an algebraic first integral. Moreover,   a lower
bound on the degree of rational first integrals can be obtained.

The rest of this article is organized as follows. In Section
\ref{s2}, we define the notion of singularity and singularity
confinement for birational autonomous discrete dynamical systems.
In Section \ref{s3}, we study the local consequences of the
existence of algebraic first integral on the confinement of the
singularities in the two-dimensional case. In Section \ref{s4}, we
extend the results of Sections \ref{s3} to arbitrary dimensions.
Applications are discussed in Section \ref{s5}.


\section{FORMULATION OF THE PROBLEM}\label{s2}

In this section we introduce and illustrate the notion of
singularity confinement by considering a particular class of
two-dimensional autonomous dynamical systems. These examples are used to
motivate a formal definition valid in arbitrary dimensions.

Throughout the article the following notation will be used: If
${\bf{g}}$ is an analytic function from $\mathbb{C}^p$ to
$\mathbb{C}^r$, then $\D {\bf{g}} (\xb)$ is its {\it{Jacobian}}
matrix evaluated at $\xb \in \mathbb{C}^p$. When
$p=r$, the \textit{Jacobian} is denoted by
$\det{\left(\D{\bf{g}}(\xb)\right)}$. If $f$ is a complex-valued
rational function on $\mathbb{C}^p$, then ${\mathrm{num}}(f)$ and
${\mathrm{den}}(f)$ denote respectively the numerator and
denominator of $f$.

\subsection{Simple examples}

The basic idea of singularity confinement is to consider the
properties of solutions
close to some singularities.
To illustrate this concept, consider the class of two-dimensional
complex dynamical systems
of the form
\begin{equation}
\left(
\begin{array}{c}
x_{n+1}\\
y_{n+1}
\end{array}
\right)
=
{\bf{f}}(x_n, y_{n})
=
\left(
\begin{array}{c}
g(x_n,y_n)\\
\xn
\end{array}
\right),
\label{examples}
\end{equation}
where $g$ is a complex-valued rational function on
$\mathbb{C}^2$,
\[
g(x,y)=\frac{p(x,y)}{q(x,y)},
\label{gdef}
\]
and $p$, $q$ are relatively prime polynomials. Two types of
singularities can be distinguished for such systems. The
singularities of  {\it{first type}} are the roots of $q$ in
$\mathbb{C}^2$. That is the values $(x,y)$ at which the vector
field ${\bf{f}}(x,y)$ in (\ref{examples}) has a singularity. The
singularities of {\it{second type}} are the points where the
Jacobian of $\bf{f}$ is zero (this type of singularities was
first considered in \cite{7b}). As an example, consider the following
discrete dynamical system:
\begin{equation}
\left(
\begin{array}{c}
x_{n+1}\\
y_{n+1}
\end{array}
\right) = \f(x_n, y_{n}) = \left(
\begin{array}{c}
\displaystyle{-x_{n}-y_{n}+a+\frac{b}{x_n}}\\
\displaystyle{\xn}
\end{array}
\right), \label{3.0.1}
\end{equation}
where $a$ and $b$ are complex numbers. The analysis is performed for
generic values of the parameters $a$ and $b$. Clearly the
right-hand-side of (\ref{3.0.1})
presents a singularity if $x_n=0$. To study this singularity we
study the behavior of nearby solutions by introducing a small
complex perturbation $\epsilon$, with $|\epsilon | > 0$. We then
set $y_0$ to be any complex number and $x_0=\epsilon$. The iterates
$x_p$ and $y_p$ are then determined for any positive nonzero
integer $p$ by the discrete dynamical system (\ref{3.0.1}). After
one iteration, we have
\begin{equation}
x_{1}=\frac{b}{\epsilon} +a-y_{0}+ {\mathcal{O}}(\epsilon).
\end{equation}
In the limit $\epsilon \rightarrow 0$,\ $|x_{1}|
\rightarrow \infty$ and, using (\ref{3.0.1}) again,
\begin{eqnarray}
\label{x22} x_{2}&=&-\frac{b}{\epsilon}+y_{0}+{\mathcal{O}}(\epsilon),\\
x_{3}&=&-\epsilon + {\mathcal{O}}(\epsilon^2).
\end{eqnarray}
The limit $\epsilon \rightarrow 0$ of the next iterate $x_{4}$ is
well defined and given by $y_0$. Despite the fact that the
function $\f$ defined by (\ref{3.0.1}) is not well defined on any
point of $\mathbb{C}^2$ of the form $(0,y)$, the limit
$(x,y)\rightarrow (0,y)$ of $\f^4(x,y)$ exists and is given by
\begin{equation}
\lim_{(x,y)\rightarrow (0,y)} \f^4(x,y)= \left(
\begin{array}{c}
y \\
0
\end{array}
\right).
\label{limf4}
\end{equation}
Moreover, since the value of the limit (\ref{limf4}) depends on
$y$, the same limit applied on the Jacobian of $\f^4$ is nonzero.
The singularity is thus ``confined'' between the iterates $0$ and
$4$ and does not propagate further. This property of confinement
is very particular and we show in this paper that it is closely
related the existence of a first integral for the system. A first
integral for a discrete dynamical system of the form
(\ref{examples}) is defined to be an analytic non-constant
complex-valued function $I$ defined almost everywhere on
$\mathbb{C}^2$ which is preserved by ${\bf{f}}$, that is
\begin{equation}
I({\bf{f}}(\xb))=I(\xb),
\label{exampleinv}
\end{equation}
for every $\xb\in \mathbb{C}^2$ for which  (\ref{exampleinv}) makes sense.
It is important to note that  if ${\bf{f}}$ is not defined
at a point $\xb^*$ where  $I$ is well-defined, then the value of $I$ at $\xb^*$
is still preserved under ${\bf{f}}$ in the limit sense, {\it{i.e.}}
\begin{equation}
\lim_{\xb\rightarrow \xb^*} I\left({\bf{f}}(\xb)\right) =I(\xb^*).
\label{limitsense}
\end{equation}
This is a consequence of the fact that ${\bf{f}}$ and $I$ are
continuous and defined almost everywhere in $\mathbb{C}^2$. In the
particular case of system (\ref{3.0.1}), one can check that the
following polynomial
\begin{equation}
{I}(x,y)=xy(x+y-a)-b(x+y)
\end{equation}
is a first integral. Moreover, following (\ref{limitsense}), the
first integral $I$ is preserved in the limit sense through the
iterations of the singularity of (\ref{3.0.1}). That is,
\begin{equation}
-by=I(0,y)=\lim_{(x,y)\rightarrow (0,y)} I\circ
\left(\f^i(x,y)\right),\;\; i=1,2,3,4.
\end{equation}
Note that most  integrable systems used in this paper are
particular cases of the so-called QRT-family of mappings
\cite{QRT89,12a}.

The confinement property for the solutions of (\ref{examples})
does not hold in general. Although there is no formal result
establishing how rare this confinement property might be, the fact
that it does not hold for most systems of the form
(\ref{examples}) has been  well established  in the literature
\cite{7,7a,9,10,11,12}.


The following system is an example where singularities are not
confined:
\begin{equation}
\label{examplenc} \left(
\begin{array}{c}
x_{n+1}\\
y_{n+1}
\end{array}
\right) = \f(\xn,\yn)= \left(
\begin{array}{c}
\displaystyle{-x_n-y_n+a+\frac{b}{x_n^3}}\\
\xn
\end{array}
\right).
\end{equation}
Consider the singularity $(x^*,y^*)=(0,y)$ and, as before, introduce
$x_0=\epsilon$ and $y_0 \in \mathbb{C}$ nonvanishing. The next iterates are
$x_1=b/\epsilon^3+(a-y_0)-\epsilon$, $x_2=-b/\epsilon^3+y_0+
\mathcal{O}({\epsilon^5})$, and
$x_3=\epsilon+\mathcal{O}({\epsilon^5})$. Further, we find that,
unlike the previous example, the appropriate  cancellations allowing
for confinement
do not occur and the next iterates are also diverging at $\epsilon=0$,
$x_4=2b/\epsilon^3+(a-y_0)-\epsilon+\mathcal{O}({\epsilon^5})$,
and $x_5= -2b/\epsilon^3+y_0+ \mathcal{O}({\epsilon^5})$.
In general, the sequence of limits $\epsilon \rightarrow 0$ of the $|x_p|$ for
$p>0$ repeats the formal triplet ($\infty$, $\infty$, $0$)
indefinitely. This assertion is proven by induction based on the equality
\begin{equation}
\f^3\left( \begin{array}{l} -nb/\epsilon^3+y_0+
\mathcal{O}({\epsilon^5}) \\
\\
nb/\epsilon^3+(a-y_0)- \epsilon+\mathcal{O}({\epsilon^5})
\end{array}
\right)
= \left(
\begin{array}{l}
-(n+1)b/\epsilon^3+y_0+ \mathcal{O}({\epsilon^5})\\
\\
(n+1)b/\epsilon^3+(a-y_0)- \epsilon+\mathcal{O}({\epsilon^5})
\end{array}
\right)\label{nonconfinement}
\end{equation}
for any positive integer $n$.
    Therefore, the singularity is  not confined.
It is generally believed that  systems which lack the confinement
property will not be
integrable. However, there is no definite result attached to this
belief. We prove in the
next sections that we can use the analysis of the dynamics nearby the
singularities can be
used to conclude that  (\ref{examplenc}) does not admit an algebraic
first integral.

The following example illustrates the occurrence of the second
type of singularity. Consider
\begin{equation}
\left(
\begin{array}{c}
x_{n+1}\\
y_{n+1}
\end{array}
\right) =\f(\xn,\yn)= \left(
\begin{array}{c}
\displaystyle{\frac{\xn-1}{\yn}+2\xn}\\
\displaystyle{\xn}
\end{array}
\right).
\label{facileexample}
\end{equation}
At first sight, the point $(\xn,0)$ seems to be the only
singularity. However, the Jacobian of ${\f}$ in
(\ref{facileexample}) vanishes whenever $\xn =1$. If $x_0=1$ and
$y_0\neq 0$, the next iterate as determined by
(\ref{facileexample}) is given by $x_1=2$. We then have $x_2=5$,
$x_3=12$. It is not difficult to see that $x_i$ grows with respect
to $i$ and that all iterates,  of $(1,y)$ under
(\ref{facileexample}) with $y\neq 0$ are independent of $y$. The
singularity here only appears in the inverse of the Jacobian
matrix and  is not confined.

In the light of the previous examples, we  can define the
confinement property for singularities of discrete dynamical systems
of the form
(\ref{examples}). A general definition will be given in the next section. A
{\it{singularity}} of the dynamical system (\ref{examples}) is
defined to be any
point $(x^*,y^*)$ in
$\mathbb{C}^2$ at which the right-hand-side of (\ref{examples}) is
undefined or at which the Jacobian of $\f$ is zero. A singularity
$(x^*,y^*)$ is said to be {\it{confined}} if there exists a
positive integer $N$ such that both limits
\begin{equation}
\lim_{(x,y)\rightarrow (x^*,y^*)} {\bf{f}}^N(x,y),\;\;
\lim_{(x,y)\rightarrow (x^*,y^*)} \det{\left(\D{{\bf{f}}}^N
(x,y)\right)} \label{limits}
\end{equation}
exist and the second limit is nonzero.
The smallest
number
$N$ having this property is referred to as the \textit{confinement
number}, that is the number of
steps necessary for confinement.

The next example  illustrates the case when a singularity of the
second type is confined.
\begin{equation}
\left(
\begin{array}{c}
x_{n+1}\\
y_{n+1}
\end{array}
\right) =\f(\xn,\yn)= \left(
\begin{array}{c}
\displaystyle{\frac{(\xn-1/a)(\xn -a)}{\yn (\xn -b)(\xn -1/b)}}\\
\displaystyle{\xn}
\end{array}
\right),
\label{dexample}
\end{equation}
where $a$ and $b$ are two nonzero distinct complex numbers which
are also both distinct from $1/a$ and $1/b$. The right-hand side
(RHS) of (\ref{dexample}) presents a singularity of the second
type if $\xn =a$ because the Jacobian of $\f$ vanishes. If $x_0=a$
and $y_0 \neq 0$, the next iterate as determined by
(\ref{dexample}) will be $x_1=0$, then $x_{2}=1/a$ and $x_{3}$
takes an indeterminate form $0/0$. Again, we  introduce a small
complex perturbation $\epsilon$, $|\epsilon|>0$, then set $x_0 =
\epsilon$ to find that when $\epsilon \rightarrow 0$,
\begin{equation}
x_{3}\rightarrow y_0+a-\frac{1}{a},
\end{equation}
and we conclude that the singularity is confined. The dynamical
system (\ref{dexample})  has another singularity of the first type
if $x_n=b$. Applying the same procedure as before with a
perturbation $\epsilon$, one finds that the limit
$\epsilon\rightarrow 0$ of $|x_1|$, $x_2$ and $x_3$ gives,
respectively, $\infty$, $1/b$ and $by_0/(b+y_0(1-b^2))$. Thus, any
singularity of the form $(b,y_0)$ with $y_0\neq b/(b^2-1)$ is
confined in 3 steps. System (\ref{dexample}) admits the
following rational first integral
\begin{equation}
I(x,y)=\left(b+\frac{1}{b}\right)(x+y)+
\left(a+\frac{1}{a}\right)\left(\frac{1}{x}+\frac{1}{y}\right)-
\frac{(1+x^2)(1+y^2)}{xy}. \label{integralfordexample}
\end{equation}

The following example shows that the existence of a first integral
does not imply that all singularities are confined. The system
\begin{equation}
\left(
\begin{array}{c}
x_{n+1}\\
y_{n+1}
\end{array}
\right) = \f(x_n, y_{n}) = \left(
\begin{array}{c}
\displaystyle{{\xn^2}/{\yn}}\\
\displaystyle{\xn}
\end{array}
\right),
\label{exetrange}
\end{equation}
has the following first integral
\begin{equation}
I(x,y)=\frac{x\,y}{x^2+y^2}. \label{firstintegraletrange}
\end{equation}
The RHS of (\ref{exetrange}) has a pole of order 1 at any point of
$\mathbb{C}^2$ of the form $(x,0)$ with $x\neq 0$. As before, if
we study this singularity by perturbation we conclude that there
is no confinement since the Laurent expansions of the iterations
of $(x,\epsilon)$  under the discrete system (\ref{exetrange})
around $\epsilon=0$ are all divergent. Although this singularity
is not confined, one notices that any point of the form $(x,0)$
with $x\neq 0$ does not have a unique preimage under
(\ref{exetrange}). Indeed, the expression defining $\f^{-1}$ given
by
\begin{equation}
\f^{-1}(x, y) = \left(
\begin{array}{c}
\displaystyle{y}\\ \\
\displaystyle{{y^2}/{x}}
\end{array}
\right)
\label{exetrangem1}
\end{equation}
has a singularity of second type at $(x,0)$ and all the points of
this form are mapped to $(0,0)$ under ${\f}^{-1}$. This example
shows that if one wishes to establish a relationship between
singularity confinement and first integrals, particular care
should be given in distinguishing different singularities.

\subsection{Preliminaries}

The notion of singularity confinement will now be defined in a
more general setting. We consider $p$-dimensional autonomous
discrete dynamical systems described by birational mappings on
$\mathbb{C}^p$. A {\it{birational mapping}} on $\mathbb{C}^p$ is a
$\mathbb{C}^p$-valued rational function which is one-to-one almost
everywhere in $\mathbb{C}^p$ and whose inverse, where it exists,
is also represented by a rational function. The corresponding
discrete dynamical
systems associated are written:
\begin{equation}
\label{mapping}
\xbu={\bf{f}}{(\xbn)},\;\;\xbn \in \mathbb{C}^p,\;\;n\in \mathbb{Z},
\end{equation}
where ${\bf{f}}$ is a birational mapping on  $\mathbb{C}^p$ with
no explicit dependence on $n$. The $i$th {\it{iterate}} of
${\bf{x}}_0\in \mathbb{C}^p$ under (\ref{mapping}) is said to
\textit{exist} if the following limit
\begin{equation}
\lim_{\xb\rightarrow \xb_0} \f^i (\xb)
\end{equation}
exists. The $i$th
iterate is then defined to be that limit and is
denoted ${\bf{F}}^i(\xb_0)$.

\vspace{.5cm}

{\noindent}{\bf Definition.} {\it{A {\bf{singularity of type I}}
for the discrete dynamical system $\xbu={\bf{f}}{(\xbn)}$ is a point
$\xb^* \in \mathbb{C}^p$ at which ${\bf{f}}$ is not defined. The
set of singularity of type I associated with the mapping
${\bf{f}}$ in $\mathbb{C}^p$
is denoted $\SI$.}}\\
\\
{\noindent}{\bf Definition.} {\it{ A {\bf{singularity of type II}}
for the discrete dynamical system $\xbu={\bf{f}}{(\xbn)}$ is a point
$\xb^* \in \mathbb{C}^p$ such that
$\det{\left(\D{{\bf{f}}}(\xb^*)\right)}=0$. The set of singularity
of type II in $\mathbb{C}^p$
is denoted $\SII$.}}\\
\\
In the particular case of birational mappings, we have
\begin{eqnarray}
\SI&=&\left\{\xb^*\in \mathbb{C}^p \Big{|}
\mathrm{den}{({{f}}_i)}(\xb^*)=0,\;\mathrm{for\;\; some}\;\;i\in
\{1,2,\dots,p \} \right\},\\
\nonumber \\
\SII&=&\left\{\xb^*\in \mathbb{C}^p \Big{|}
\det{\left(\D{\bf{f}}(\xb^*)\right)} =0 \right\}.
\label{S1}
\end{eqnarray}
We call the sets $\SI$ and $\SII$, the {\it{singular sets of first and second
type}}, respectively. The {\it{singular set}} $S(\f)$ is defined as the
union of both sets of first and second type
\begin{equation}
S(\f)=\SI \cup \SII.
\end{equation}
For both types of singularity we can define the property of
confinement.\\
\\
{\noindent}{\bf Definition.} {\it{Let $\xb^* \in \mathbb{C}^p$ be a
singularity of type I or II for the  system $\xbu={\bf{f}}{(\xbn)}$.
The singularity
is said to be {\bf{confined}} if, for some $i_c \in \mathbb{N}^{>0}$,
the iterate
${\bf{F}}^{i_c}(\xb^*)$ exists and $\lim_{\xb\rightarrow \xb^*}
\det{\left(\D{\bf{f}}^{i_c}(\xb)\right)}\neq
0$.}} \\
\\
The lowest such $i_c$
is referred to as the {\it{confinement number}} that is the number of steps
necessary for confinement. From now on, we assume that ${\bf f}$ is a
birational
mapping.  The following lemma proves to be useful.\\
\\
{\noindent}{\bf Lemma 2.1.} {\it{The image of $\SII$ under the
mapping ${\bf f}$ lies inside an algebraic
variety of codimension 2.}}\\
\\
That is, the birational mapping ${\bf f}$ is one-to-one
only on the
subset of $\mathbb{C}^p$ defined by
\begin{equation}
O_{\bf f}\equiv \mathbb{C}^p \backslash S(\f).
\end{equation}
As a consequence, the function ${\bf f}$ restricted to $O_{\bf f}$  defines a
rational diffeomorphism
\begin{equation}
{\bf f}:O_{\bf f} \rightarrow O_{\bf f}'\equiv {\bf f} (O_{\bf f}),
\end{equation}
whose inverse is also a rational function.

\begin{proof}
Assume that $\SII \neq \emptyset$. Let $\xb_0 \in \SII$ and let
$\xb_1 = {\bf{f}}(\xb_0)$.
   From the inverse function theorem, it follows that
\begin{equation}
\lim_{{\xb\rightarrow \xb_1}}
     \left|\left|\;
\det{\left(
\D\,\f^{-1}(\xb)
\right)}\;
\right|\right| = \infty.
\label{inf}
\end{equation}
Thus, for some $i$'s in $\{1,2,\dots,p\}$, the denominator of
$({\bf{f}}^{-1})_i$ evaluated at $\xb_1$ is zero. Denote those $i$
by $i_k,\;\;k=1,2,\dots,r$. We now prove that at least for one
$i_k$, ${\mathrm{num}}\left({(\bf{f}}^{-1})_{i_k}\right)$ also
vanishes at $\xb_1$. By contradiction, suppose that none of these
numerators vanishes at $\xb_1$. Thus
\begin{equation}
\lim_{\xb\rightarrow \xb_1} \|\f^{-1}(\xb)\|=\infty.
\label{inf1}
\end{equation}
On the other hand, we have
\begin{equation}
\lim_{\xb\rightarrow \xb_0} \|\f^{-1}({\bf{f}}(\xb))\|=\xb_0,
\end{equation}
which contradicts equation (\ref{inf1}). Thus,
    $\xb_1$ must be in one
of the following
sets
\begin{equation}
A_i=\{\xb \in \mathbb{C}^p
|\;\;\mathrm{num}((\f^{-1})_i)(\xb)=\mathrm{den}((\f^{-1})_i)(\xb)=0\},
\;\;\;\;i\in\{1,2,\dots,p\}.
\end{equation}
This concludes the proof since
each set $A_i$ is defined as the intersection
of the zero level sets of two relatively prime polynomials.
\end{proof}

As mentioned earlier, the singularity confinement
turns out to be closely related to the existence of a first
integral. This relationship will be explored in the next section. We
   define  the notion of a first integral.\\
\\
{\noindent}{\bf Definition.} {\it{A {\bf{first integral}} of
$\xbu={\bf{f}}{(\xbn)}$
is an analytic function
$I:U\subset \mathbb{C}^p \rightarrow \mathbb{C}$ where $U$ is a
dense subset of $\mathbb{C}^p$ such that
\begin{itemize}
\item[(a)]
$\left|\left|\left(\frac{\partial{I}}{\partial{x_1}},\frac{\partial{I}}{\partial{x_2}},\ldots,
\frac{\partial{I}}{\partial{x_p}}\right)\right|\right|\neq
0$ almost everywhere in $U$.
\item[(b)]
$I({\bf{f}}(\xb))=I(\xb)$.
\end{itemize}}}

\vspace{.3cm}

{\noindent}A first integral can exist in several forms.
\\
\\
{\noindent}{\bf Definition.} {\it{ A first integral $I$ of
$\xbu={\bf{f}}{(\xbn)}$ is
{\bf{polynomial}} (resp. {\bf{rational}}) if the function $I({\bf{x}})$ is a
polynomial (resp. rational) function of $\xb \in \mathbb{C}^p$. A
complex-valued function $f({\bf{x}})$ is {\bf{algebraic}} over
$\mathbb{C}$ if there exist $s>0$ and $q_0,\dots,q_s$ rational in
${\xb}$ such that
\begin{equation}
q_0+q_1 f+\dots+q_s f^s=0.
\label{algebraic}
\end{equation}
If $s$ is the smallest positive integer such that
(\ref{algebraic}) holds, the relation (\ref{algebraic}) is
referred to as the minimal polynomial of $f$. A first integral $I$
is {\bf{algebraic}} if the function
$I({\bf{x}})$ is an algebraic function.}}\\
\\
Many notions of integrability are used in the literature. In this
paper, we concentrate on algebraic integrability. In order to
introduce the latter, we are interested in determining
precisely ``how many'' algebraic first integrals a given discrete
system must admit to be algebraically integrable. To
further clarify this problem, suppose that the algebraic functions
$A_1({\bf{x}}), A_2({\bf{x}}),\ldots, A_r({\bf{x}})$ defined on
$\mathbb{C}^p$ are all first integrals for the system
(\ref{mapping}) and $F(z_1,z_2,\ldots,z_r)$ is an algebraic function
on $\mathbb{C}^r$. Then, it is clear that
$A({\bf{x}})=F(A_1({\bf{x}}),A_2({\bf{x}}),\ldots, A_r({\bf{x}}))$ is
also a first integral. However, this  new first integral adds no
knowledge to the given problem since it depends on the
other first integrals.\\

{\noindent}{\bf{Definition.}} {\it{Let $A_1({\bf{x}}),
A_2({\bf{x}}),\ldots, A_r({\bf{x}})$ be smooth complex-valued
functions defined on
    a domain $D\subset \mathbb{C}^p$. Then
\begin{itemize}
\item[(a)] $A_1, A_2,\ldots, A_r$ are  {\bf{functionally dependent}} if,
for each $\xb\in D$,
there is a neighborhood $U$ of $\xb$ and a
    smooth complex-valued function $F(z_1,z_2,\ldots,z_r)$ not identically
zero on any subset
of $\mathbb{C}^p$
    such that
\begin{equation}
F(A_1({\bf{x}}),A_2({\bf{x}}),\ldots,
A_r({\bf{x}}))=0.
\end{equation}
\item[(b)] $A_1, A_2,\ldots, A_r$ are  {\bf functionally independent} if
they are not functionally
dependent when restricted to any open subset of $D$.
\end{itemize}
}}

A simple way to determine if a set of functions are functionally
independent is given
by the following theorem (see for example \cite{Olver}).\\
\\
{\noindent}{\bf{Theorem.}} {\it{
Let $A=(A_1,A_2,\ldots,A_r)$ be a set
of smooth functions from a domain $D\in\mathbb{C}^p$ to
$\mathbb{C}^r$. Then,  $A_1({\bf{x}}), A_2({\bf{x}}),\ldots,
A_r({\bf{x}})$ are functionally dependent if and only if
$\D{{\bf{A}}} (\xb^*)$ has
rank strictly less than $r$ for all $\xb^*\in D$.}} \\
\\
We introduce the notion of integrability
considered in this paper.\\
\\
{\noindent}{\bf Definition.}{\it{ The discrete dynamical system
$\xbu={\bf{f}}{(\xbn)}$  is said to be {\bf{algebraically integrable}} if
it admits $(p-1)$
algebraic functionally independent first integrals.}}\\
\\
The following theorem is a direct extension of Brun's
\cite{Goriely02,Kummer90} theorem to discrete systems and is useful
to
extend the results we obtain for rational first integrals to algebraic ones.\\
\\
{\noindent}{\bf Theorem.}: {\it{If a discrete dynamical system of
$\xbu={\bf{f}}{(\xbn)}$  has $k$ functionally independent
algebraic first integrals, then it has $k$ functionally
independent rational first integrals.}}
\begin{proof}
Let $I$ be an algebraic first integral and let
\begin{equation}
P(I)=q_0+q_1 I +\dots+ q_{s-1} I^{s-1}+I^s
\label{minimal}
\end{equation}
be its minimal polynomial, where $q_i$ is a rational function of
$\xb$. Since $I$
depends non trivially
on $\xb$, let $i$ be such that $q_i$ is a non constant rational
function of $\xb$.
Since $I(\f(\xb))=I(\xb)$,
we have that
\begin{equation}
q_0(\f(\xb))-q_0(\xb)+(q_1(\f(\xb))-q_1(\xb)) I(\xb) +\dots+
(q_{s-1}(\f(\xb))-q_{s-1}(\xb)) I^{s-1}(\xb)=0.
\end{equation}
Since $P$ is minimal, we have $q_i(\f(\xb)))=q_i(\xb)$ and
each $q_i$ is a first integral. Now, let $I'$ be another
independent first integral whose minimal polynomial is
\begin{equation}
P(I')=q_0'+q_1' I' +\dots+ q_{s'-1}' {I'}^{s'-1}+{I'}^{s'}.
\label{minimal'}
\end{equation}
The independence between the two first integrals implies that
there exist $i<s,\;i'<s'$ such that
$q_i,\;q_{i'}$ are two
independent non constant rational first integrals. By induction, one
can build $k$ independent rational first integrals.
\end{proof}

Therefore, rational integrability implies algebraic integrability and
it is  sufficient to consider
discrete dynamical systems of the
form (\ref{mapping})
which admit $(p-1)$ functionally independent rational first integrals
$R_i(\xb)=P_i(\xb)/Q_i(\xb)$
where $i=1,2,\dots,p-1$ and
$P_i$ and $Q_i$
are relatively prime polynomials. Let $L_{i,c}$ be the level set of
the first integral
$R_i$ corresponding to the value
$c_i\in \mathbb{C}$. Since  $\f$ is continuous, it  leaves
invariant the
closure of the
level set given by
\begin{equation}
\overline{L}_{i,c}=\{\xb \in \mathbb{C}^p | P_i(\xb) -c\, Q_i(\xb)=0 \}.
\label{Lci}
\end{equation}


\section{The two-dimensional case}\label{s3}

In this section we study the local implications of algebraic
integrability on the confinement of  singularities for
two-dimensional discrete dynamical systems.
We consider a discrete dynamical systems of the form
(\ref{mapping}) which is algebraically integrable, that is, it
possesses a non trivial rational first integral
$R(\xb)=P(\xb)/Q(\xb)$ where $P$ and $Q$ are some polynomials over
$\mathbb{C}^2$. Let $L_{c}$ be the level set of $R$ corresponding
to the value $c\in \mathbb{C}$. Then, as noted before, $\f$ leaves
invariant the closure of the level set $\Lc$.

We first focus our attention on singularities of the first type.
The singular set of first type $\SI$ is an algebraic variety since it is
defined as the set of zeros of polynomials. We consider
its irreducible decomposition
\begin{equation}
\SI=\bigcup_{i=1}^{d} S_I^{(i)}({\bf{f}})
\label{s1decomposition}
\end{equation}
where each irreducible component $S_I^{(i)}({\bf{f}})$ is defined as the
zero set of an irreducible polynomial on $\mathbb{C}^2$.

In Example (\ref{exetrange}), we showed that the
existence of a rational first integral does not ensure that all
singularities are confined. However,  in this case
   it is not possible to ``enter'' the singularity, meaning that
it does not have a unique preimage under the mapping. Moreover, we
   know that confinement is a generic property, that is, it is
only concerned with  dense subsets of irreducible components of $S_I$.
For instance, in Example (\ref{dexample})  all singularities of type I of
the form $(b,y_0)$ confine in 3 steps except a particular singularity defined
by $y_0 = b/(b^2-1)$.
We state the general theorem relating the
confinement of singularities to the existence of a rational first
integral.\\
\\
{\noindent}{\bf{Theorem 3.1.}} {\it{Consider a two-dimensional
birational mapping $\xbu={\bf{f}}{(\xbn)}$ and assume
it has a rational first integral $R(\xb)=P(\xb)/Q(\xb)$. Then, for
each irreducible component $S_I^{(j)}({\bf{f}})$, we have either
\begin{itemize}
\item[(a)]
There exists $k_j\in \mathbb{N}^{>0}$ such that
\begin{equation}
\label{finitude}
\lim_{\xb\rightarrow \xb^*}{\bf{f}}^{k_j}(\xb)
\end{equation}
exists for almost all $\xb^*\in S_I^{(j)}({\bf{f}})$, or,
\item[(b)]
There exists $l_j \in \mathbb{N}^{>0}$ such that
$S_I^{(j)}({\bf{f}})$ lies inside the singular set of the birational
mapping defining ${\bf{f}}^{-l_j}$.
\end{itemize}
}}
\begin{proof}
Without loss of generality, take $j=1$.
Consider two cases:
\begin{itemize}
\item[1)]
$S_I^{(1)}({\bf{f}})\not\subset \Lc$ for all $c\in \mathbb{C}$.

Let $\xb^*\in S_I^{(1)}({\bf{f}})$ such that
$\lim_{\xb\rightarrow\xb^*}||f_i(\xb)||$
exists or is infinity (as opposed to undefined) for $i=1,2$. Suppose
that $\xb^*\in\Lcs$ for only one
$c^*\in\mathbb{C}$ and that
there is a neighborhood of $\xb^*$ on $\Lcs$
whose only intersection with any of the singular sets
associated with the mappings $\f^i$, $i=1,2,\ldots,q+1$ ($q$ is defined
below) is $\xb^*$ itself.
The set of points satisfying
the above
properties is dense in $S_I^{(1)}({\bf{f}})$.
Let $q$ be the number of paths
${\bf{p}}_i(\epsilon)$, $i=1,2,..,q$
in $\mathbb{C}^2$
such that
\begin{itemize}
\item[(i)]
${\bf{p}}_i(\epsilon)\in \Lcs$ for $\epsilon$ small enough. That is,
\begin{equation}
P({\bf{p}}_i(\epsilon))-c^*\,Q({\bf{p}}_i(\epsilon))=0.
\end{equation}
\item[(ii)]
$||{\bf{p}}_i(\epsilon)||\rightarrow \infty$ as $\epsilon\rightarrow 0$
and at least one of the components of ${\bf{p}}_i(\epsilon)$
    is of the form
$1/\epsilon$.
\end{itemize}
Note that the number $q$ can depend on the complex number $c^*$
but the values it can take for different $c^*$ can be bounded
above by a number depending only on the degrees of $P$ and $Q$.

Now, consider another path $\mathbf{x}(\sigma)\in \Lcs$ $\forall
\sigma$ such that
\begin{equation}
\lim_{\sigma\rightarrow 0} \xb(\sigma)=\xb^*.
\end{equation}
Since there
is a neighborhood of $\xb^*$ on $\Lcs$ whose only intersection with
$S({\bf{f}})$ is $\xb^*$, $\f(\xb(\sigma))$
is well-defined for $\sigma$ nonzero and small enough. Moreover,
since $\f$ preserves $\Lcs$, ${\bf{f}}(\sigma)$ also lies in $\Lcs$, and
\begin{equation}
\lim_{\sigma\rightarrow 0} ||\f(\xb(\sigma))||=\infty.
\end{equation}
Therefore, there exists a change of variable
$\sigma=\sigma(\epsilon)$ of order ${\cal{O}}(\epsilon^s)$ for some
$s>0$, so that
$\f(\xb(\sigma))$ is
one of the paths ${\bf{p}}_i$ defined above:
\begin{equation}
\f(\xb(\sigma(\epsilon)))={\bf{p}}_i(\epsilon)
\end{equation}
for some $1\leq i\leq q$. Our claim is that there exists $m$ with
$m\leq q+1$ such
that the limit as $\sigma\rightarrow 0$ of $\f^m(\xb(\sigma))$ exists
and is finite.
Suppose by contradiction
that the iterates
$\f^k(\xb(\sigma))$
$k=1,2,3,\ldots,q+1$ are all
divergent at $\sigma=0$. Each iterate $\f^k(\xb(\sigma))$ is
associated with a unique path
${\bf{p}}_{i_k}(\epsilon)$. This implies that
the $k^{\mathrm{th}}$ preimage of ${\bf{p}}_{i_k}(\epsilon)$
is finite in the limit $\epsilon\rightarrow 0$
and no other $\f^l(\xb(\sigma))$ for
$1\leq k < l\leq q+1$  corresponds to the same
    ${\bf{p}}_{i_k}(\epsilon)$. Hence the relation between
the $q+1$ iterates that are divergent at $\sigma=0$ and the paths
${\bf{p}}_{i}(\epsilon)$
is injective. This is a contradiction since there are only
$q$  paths ${\bf{p}}_{i}(\epsilon)$.

\item[2)] There exists a $c\in \mathbb{C}$ such that
$S_I^{(1)}({\bf{f}})\subset\Lc$.
Let
\begin{equation}
\Lc=\bigcup_{i=1}^{r}\Lc^{(i)}
\end{equation}
be
the irreducible decomposition of $\Lc$ for any $c\in \mathbb{C}$.
If $S_I^{(1)}(\f)$ lies inside the singular set of $\f^{-1}$, then
the theorem is
proved. Otherwise, let $k\in \mathbb{N}^{>0}$ be such that $S_I^{(1)}(\f)$
does not lie in a
singular set associated with the mappings $\f^{-m}$, $m=1,2,\ldots,k$.
Then,
\begin{equation}
\overline{\f^{-m}(S_I^{(1)}(\f))}=\bigcup_{i\in
B_m}\Lc^{(i)},\;\;\;\;1\leq l\leq k
\label{inverse}
\end{equation}
where each $B_m$ is a subset of $\{1,2,\ldots,r\}$. The $k$ sets
in (\ref{inverse}) are distinct, otherwise, by contradiction
there exist $m_1,m_2 \in \mathbb{N}^{>0}$ such that
\begin{equation}
\overline{\f^{-m_1}(S_I^{(1)}(\f))}=
\overline{\f^{-m_2}(S_I^{(1)}(\f))},\;\;\;\;1\leq m_1< m_2\leq k.
\end{equation}
This implies that
\begin{equation}
S_I^{(1)}(\f)=\overline{\f^{-1}\left(\f^{(m_1-m_2+1)}(S_I^{(1)}(\f))\right)}
\end{equation}
which is a contradiction since $\f$ is undefined
almost everywhere on $S_I^{(1)}(\f)$. Hence the $k$
sets in (\ref{inverse}) are distinct. However, there are only finitely
many possibilities for the sets $B_m$ in (\ref{inverse}). Hence
    $S_I^{(1)}(\f)$ must lie inside the singular set of $\f^{-l_1}$ for
some $l_1$.
\end{itemize}
\end{proof}

Notice that the condition (\ref{finitude}) does not mean that the singularity
is confined since the condition that the Jacobian be nonzero
must be satisfied. In order to solve this problem, the singularities
of type II have to
be considered. The singular set of second type is not, in general, an
algebraic variety
but its closure
\begin{equation}
\overline{\SII}=\{\xb^*\in \mathbb{C}^p \Big{|}
{\mathrm{num}}\left(\det{\left(\D{{\bf{f}}}
(\xb^*)\right)}\right) =0 \}
\end{equation}
is. Hence, it is possible to introduce the irreducible decomposition
\begin{equation}
\overline{\SII}=\bigcup_{i=1}^{s} \overline{\SIIw^{(i)}(\f)},
\label{s2decomposition}
\end{equation}
where each $\SIIw^{(i)}(\f)$ is a subset of $\SII$.
It is now possible to state the general theorem for singularities of
second type.\\
\\
{\noindent}{\bf{Theorem 3.2.}} {\it{Consider a two-dimensional
birational mapping $\xbu={\bf{f}}{(\xbn)}$  and assume
it has a rational first integral $R(\xb)=P(\xb)/Q(\xb)$. Then, for each
irreducible component $\overline{\SIIw^{(j)}({\bf{f}})}$, we have either
\begin{itemize}
\item[(a)]
There exists $k_j\in \mathbb{N}^{>0}$ such that
\begin{equation}
\label{nonzeroude}
\lim_{\xb\rightarrow \xb^*}\det{\left(\D{{\bf{f^{k_j}}}}(\xb)\right)}
\end{equation}
exists and is nonzero for almost all $\xb\in {\SIIw^{(j)}({\bf{f}})}$, or,
\item[(b)]
There exists $l_j \in \mathbb{N}^{>0}$ such that
${\SIIw^{(j)}({\bf{f}})}$ lies inside the singular set of ${\bf{f}}^{-l_j}$.
\end{itemize}
}}
\begin{proof}
The proof is similar to the proof of Theorem 3.1.
Again, without loss of generality, we take $j=1$ and
consider two cases.
\begin{itemize}
\item[1)]
$\overline{\SIIw^{(1)}({\bf{f}})}\not\subset \Lc$ for all $c\in \mathbb{C}$.

The image of $\SIIw^{(1)}({\bf{f}})$ under the mapping $\f$
is a finite subset of $\mathbb{C}^2$, thanks to Lemma 2.1. Moreover,
$\SIIw^{(1)}({\bf{f}})$ is not contained inside any level set. For any point
in the image of $\SIIw^{(1)}({\bf{f}})$ it is therefore
possible to choose two distinct preimages in $\SIIw^{(1)}({\bf{f}})$
which lie in two different level sets.
Since $\f$
preserves $\Lc$, every
point in the image of $\SIIw^{(1)}({\bf{f}})$ must lie in the set
\begin{equation}
\bigcap_{c\in\mathbb{C}} \Lc=\{\xb\in\mathbb{C}^2 \Big{|} P(\xb)=Q(\xb)=0\}.
\end{equation}
The elements of this finite set are denoted $\s_i$, $i=1,2,\ldots,m$.
Note that in the case of a polynomial first integral, $Q(\xb)=1$
and this set is empty.

Now, choose $\xb^*\in \SIIw^{(1)}({\bf{f}})$
such that $\xb^*\in\Lcs$ for only one
$c^*\in\mathbb{C}$ and
there is a neighborhood of $\xb^*$ on $\Lcs$
whose only intersection with any of the singular sets
associated with the mappings $\f^i$, $i=1,2,\ldots,q+r+1$ ($q$ and $r$
are defined below) is $\xb^*$
itself. The set of points satisfying
the above
properties is dense in $\SIIw^{(1)}({\bf{f}})$.

Consider the $q$ paths
${\bf{p}}_i(\epsilon)$
described in the proof of Theorem 3.1.
Consider also the $r$ paths
${\bf{w}}_i(\epsilon)$, $i=1,2,..,r$
in $\mathbb{C}^2$
satisfying
\begin{itemize}
\item[(i)]
${\bf{w}}_i(\epsilon)\in \Lcs$ for $\epsilon$ small enough.
\item[(ii)]
$||{\bf{w}}_i(\epsilon)||\rightarrow \s_{j_i}$ as $\epsilon\rightarrow 0$
and one of the components ($k=1$ or $2$) of ${\bf{w}}_i(\epsilon)$
    is of the form
$(\s_{j_i})_k+\epsilon$ for some $1\leq j_i\leq m$.
\end{itemize}

Now, consider another path $\xb(\sigma) \in \Lcs$ such that
\begin{equation}
\lim_{\sigma\rightarrow 0} \xb(\sigma)=\xb^*.
\end{equation}
Since there
is a neighborhood of $\xb^*$ on $\Lcs$ whose only intersection with
$S({\bf{f}})$ is $\xb^*$, $\f(\xb(\sigma))$
is well-defined for $\sigma$ nonzero and small enough. Moreover
since $\f$ preserves $\Lcs$, $\f(\xb(\sigma))\in \Lcs$, and
\begin{equation}
\lim_{\sigma\rightarrow 0} \f(\xb(\sigma)) = \s_i
\end{equation}
for some $i$ such that $1\leq i\leq r$.  Following
    the proof of Theorem 3.1, $\f(\xb(\sigma))$ can be associated
with a path ${\bf{w}}_j(\epsilon)$ (whose limit as $\sigma\rightarrow
0$ is $\s_i$).
Suppose by contradiction that the $d+r+1$ iterates
$\f^k(\xb(\sigma))$
$k=1,2,\ldots,d+r+1$ are either
divergent at $\sigma=0$ or have a limit $\sigma\rightarrow 0$ equal
to one of the $\s_i$.
Each iterate $\f^{k}$ is associated with a path
${\bf{p}}_{i_k}(\epsilon)$ or ${\bf{w}}_{i_k}(\epsilon)$.
This implies that the unique $k^{\mathrm{th}}$ preimage of
${\bf{p}}_{i_k}(\epsilon)$ or
${\bf{w}}_{i_k}(\epsilon)$ must converge to $\xb^*$ as $\epsilon\rightarrow 0$.
The preimage is unique because
there is a neighborhood of $\xb^*$ on $\Lcs$
whose only intersection with any of the singular sets
associated with the mappings $\f^i$, $i=1,2,\ldots,q+r+1$ ($q$ and $r$
are defined below) is $\xb^*$
itself. Since $\xb^*$ is not one of the $\s_i$, there cannot be two iterates
$\f^{k_1}(\xb(\sigma))$ and $\f^{k_2}(\xb(\sigma))$, $k_1,k_2 \leq q+r+1$
associated with the same path.
Since there are only $q+r+1$ paths ${\bf{p}}_{i}(\epsilon)$ and
${\bf{w}}_{i}(\epsilon)$,
there
exists a $k\leq q+r+1$ such that the limit $\sigma\rightarrow 0$ of the
$k^{\mathrm{th}}$ iterate of
$\xb(\sigma)$ must be finite and not equal to one $\s_i$ as
$\sigma\rightarrow 0$.
Therefore, the limit $\sigma\rightarrow 0$ of the Jacobian of
$\f^{k_1}$ evaluated at
$\mathbf{x}(\sigma)$ must be a finite nonvanishing number. This ends
the proof in the first case.

\item[2)] There exists a $c\in \mathbb{C}$ such that
$S_I^{(1)}({\bf{f}})\subset \Lc$.
The argument leading to the conclusion that
    ${\SIIw^{(j)}(\f)}$ lies inside the singular set of $\f^{-l_1}$ for
some $l_1$
is identical to the second case considered in the proof of the
preceding theorem.
\end{itemize}
\end{proof}

With Theorem 3.2, it is now easy to see that singularities of type
I satisfying the condition (\ref{finitude}) are confined. Indeed,
suppose $k_j$ is the number considered in the first part of
Theorem 3.1. If the Jacobian of $\f^{k_j}$ is zero almost
everywhere on $S_I^{j}(\f)$, then  $S_I^{(j)}(\f)$ becomes a
subset of $\overline{S_{II}(\f^{k_j})}$. According to the proof of
Theorem 3.1 we know that $S_I^{(j)}(\f) \not\subset \Lc$ for all
$c\in \mathbb{C}$. Thus, as a consequence of Theorem 3.2, the
singularities in $S_I^{(j)}(\f)$ must be confined. We can
   state a general theorem including  both types of
singularities. To do so,  we introduce
the irreducible decomposition associated with $S(\f)$
\begin{equation}
S(\f)=\bigcup_{i=1}^{d+s} S^{(i)}(\f)
\end{equation}
where $d$ and $s$ are defined in (\ref{s1decomposition}) and
(\ref{s2decomposition}).\\
\\
{\noindent}{\bf{Theorem 3.3.}} {\it{
Consider a two-dimensional
birational mapping $\xbu={\bf{f}}{(\xbn)}$  and assume
it has a rational first integral $R(\xb)=P(\xb)/Q(\xb)$. Let
$S({\bf{f}})$ be its
singular set with irreducible components $S^{(j)}({\bf{f}})$. Then,
for each $j$,
either
\begin{itemize}
\item[(a)]
Almost all singularities of $S^{(j)}({\bf{f}})$ are confined
in the same number of steps, or,
\item[(b)]
There exists $l_j \in \mathbb{N}^{>0}$ such that
$S^{(j)}({\bf{f}})$ lies inside the singular set of ${\bf{f}}^{-l_j}$.
\end{itemize}
}}


\section{Arbitrary dimensions}\label{s4}

In this section, we extend the result of the previous section to
birational mappings of arbitrary dimension $p$. The mappings
are assumed to be algebraically integrable, that is, they admit $(p-1)$
functionally independent first integrals
$R_i(\xb)=P_i(\xb)/Q_i(\xb)$, $i=1,2,\ldots,p-1$. Then, the closure of the
level sets associated with $\c \in \mathbb{C}^{p-1}$ is given by
\begin{equation}
\Lcv=\{\xb \in \mathbb{C}^p | P_i(\xb) -c_i\,
Q_i(\xb)=0,\;\;{\mathrm{for\ }}1\leq i \leq p-1 \}=
\bigcap_{i=1}^{p-1} \overline{L}_{i,c_i},
\end{equation}
with $\overline{L}_{i,c_i}$ defined in (\ref{Lci}). A set of
constants $\c \in \mathbb{C}^{p-1}$ is
said to be {\it{regular}} for the first integrals ${\bf{R}} =
(R_1,R_2,\ldots,R_{p-1})$ if
$
\D {\bf{R}}(\xb)
$
has rank equal to $(p-1)$ almost everywhere on
${\bf{R}}^{-1}(\c)$.  The set
$\Lcv$ is then an algebraic variety of codimension $(p-1)$.

Our main Theorems 3.1, 3.2, and 3.3 can be readily generalized to
arbitrary dimensions and we only give here an outline of the proofs.\\
\\
{\noindent}{\bf{Theorem 4.1.}} {\it{Consider a $p$-dimensional
birational mapping $\xbu={\bf{f}}{(\xbn)}$ and assume
it is algebraically integrable with first integrals
$R_i(\xb)=P_i(\xb)/Q_i(\xb)$, $i=1,2,\ldots,p-1$. Then for each irreducible
component $S_I^{(j)}({\bf{f}})$, we have either
\begin{itemize}
\item[(a)]
There exists $k_j\in \mathbb{N}^{>0}$ such that
\begin{equation}
\label{finitudep}
\lim_{\xb\rightarrow \xb^*}{\bf{f}}^{k_j}(\xb)
\end{equation}
exists for almost all $\xb\in S_I^{(j)}({\bf{f}})$, or,
\item[(b)]
There exists $l_j \in \mathbb{N}^{>0}$ such that
$S_I^{(j)}({\bf{f}})$ lies inside the singular set of the birational
mapping defining ${\bf{f}}^{-l_j}$.
\end{itemize}
}}
\begin{proof}
Without loss of generality, take $j=1$.
Consider two cases:
\begin{itemize}
\item[1)]
$S_I^{(1)}({\bf{f}})\not\subset \overline{L}_{i,c}$ for all $\c\in
\mathbb{C}^{p-1}$ and $1\leq i\leq p-1$.

Let $\xb^*\in S_I^{(1)}({\bf{f}})$ such that
$\lim_{\xb\rightarrow\xb^*}||f_i(\xb)||$
exists or is infinite (as opposed to undefined) for $i=1,\ldots,p$.
Suppose that
$\xb^*\in\Lcsv$ for only one $\c^*\in \mathbb{C}^{p-1}$ which is regular
and that
there is a neighborhood of $\xb^*$ on $\Lcsv$
whose only intersection with the singular sets
associated with the mappings $\f^i$, $i=1,2,\ldots,q+1$ ($q$ is defined
below) is $\xb^*$ itself.
The set of points satisfying
the above
properties is dense in $S_I^{(1)}({\bf{f}})$.
Let $q$ be the number of paths
${\bf{p}}_i(\epsilon)$, $i=1,2,..,q$
in $\mathbb{C}^p$
such that
\begin{itemize}
\item[(i)]
${\bf{p}}_i(\epsilon)\in \Lcsv$ for $\epsilon$ small enough. That is,
\begin{equation}
P_k({\bf{p}}_i(\epsilon))-c_k^*\,Q_k({\bf{p}}_i(\epsilon))=0,\;\;k=1,2,\ldots,p-1.
\end{equation}
\item[(ii)]
$||{\bf{p}}_i(\epsilon)||\rightarrow \infty$ as $\epsilon\rightarrow 0$
and at least one of the components of ${\bf{p}}_i(\epsilon)$
    is of the form
$1/\epsilon$.
\end{itemize}

Using the fact that $\Lcsv$ is of codimension $p-1$, the rest of the
proof follows
exactly the similar case in the proof of Theorem 3.1.

\item[2)] There exists a $c\in \mathbb{C}$ and a $1\leq i\leq p-1$
such that $S_I^{(1)}({\bf{f}})\subset\overline{L}_{i,c}$.

This part of the proof is identical to the corresponding part in the
proof of Theorem 3.1 except that
$\overline{L}_{i,c}$ is considered instead of $\Lc$.
\end{itemize}
\end{proof}
%
{\noindent}{\bf{Theorem 4.2.}} {\it{ {Consider a $p$-dimensional
birational mapping $\xbu={\bf{f}}{(\xbn)}$  and assume
it is algebraically integrable with rational first integrals
$R_i(\xb)=P_i(\xb)/Q_i(\xb)$, $i=1,2,\ldots,p-1$. Then, for each
irreducible component $\overline{\SIIw^{(j)}({\bf{f}})}$, we have either
\begin{itemize}
\item[(a)]
There exists $k_j\in \mathbb{N}^{>0}$ such that
\begin{equation}
\label{nonzeroudep}
\lim_{\xb\rightarrow \xb^*}\det{\left(\D{{\bf{f^{k_j}}}}(\xb)\right)}
\end{equation}
exists and is nonzero for almost all $\xb\in {\SIIw^{(j)}({\bf{f}})}$, or,
\item[(b)]
There exists $l_j \in \mathbb{N}^{>0}$ such that
${\SIIw^{(j)}({\bf{f}})}$ lies inside the singular set of ${\bf{f}}^{-l_j}$.
\end{itemize}
}}
\begin{proof}
The proof is similar to the proof of Theorem 3.2 and the only
difference comes in the first case, when
$\overline{\SIIw^{(1)}({\bf{f}})}\not\subset\overline{L}_{i,c}$
for all $c\in \mathbb{C}$ and $1\leq i\leq p-1$. Following the
proof of Theorem 3.2, for any point in the image of
$\SIIw^{(1)}({\bf{f}})$ it is possible to choose two distinct
preimages in $\SIIw^{(1)}({\bf{f}})$ which lie in two different
level sets. Indeed, from Lemma 2.1,
    the image of
$\SIIw^{(1)}({\bf{f}})$ under the mapping $\f$
lies in a subset of codimension 2 inside $\mathbb{C}^{p}$.
Since $\f$
preserves  $\overline{L}_{i,c}$,
the points in the image of $\SIIw^{(1)}({\bf{f}})$ must lie in the set
\begin{equation}
\bigcap_{c\in\mathbb{C},i\leq p-1}
\overline{L}_{i,c}=\{\xb\in\mathbb{C}^2 \Big{|} P_i(\xb)=Q_i(\xb)=0\}.
\end{equation}
Since the first integrals are functionally independent, the set
defined above is finite.
The rest of the proof is similar to the proof of Theorem 3.2 except
that one should specify
that $\c^*$ has to be regular.
\end{proof}
%
{\noindent}{\bf{Theorem 4.3.}} {\it{
Consider a $p$-dimensional
birational mapping $\xbu={\bf{f}}{(\xbn)}$  and assume
it is algebraically integrable. Let
$S({\bf{f}})$ be its
singular set with irreducible components $S^{(j)}({\bf{f}})$. Then,
for each $j$,
either
\begin{itemize}
\item[(a)]
Almost all singularities of $S^{(j)}({\bf{f}})$ are confined
in the same number of steps, or,
\item[(b)]
There exists $l_j \in \mathbb{N}^{>0}$ such that
$S^{(j)}({\bf{f}})$ lies inside the singular set of ${\bf{f}}^{-l_j}$.
\end{itemize}
}}}

\section{Corollaries and applications}\label{s5}

In the previous two sections, the existence of first integrals was
assumed to find local information on the confinement property.
Here, the information given by the singularity confinement
property is used to obtain global information on the discrete
dynamical systems such as   the non-existence of algebraic first
integral and the degree of possible rational first integrals. The
results of this section  can be considered as a
discrete analog of Yoshida's Theorem \cite{yo83,yo83b,Goriely02}
which, for ODEs, relates the Kovalevskaya exponents  given by
the Painlev\'e test to the degree of a rational first integral.
The first corollary is a direct consequence of Theorem 4.3.\\
\\
{\noindent}{\bf{Corollary 5.1.}} {\it{Consider a $p$-dimensional
birational mapping $\xbu={\bf{f}}{(\xbn)}$ with a non-empty
singular set
$S({\bf{f}})$. If there exists an irreducible
components $S^{(j)}({\bf{f}})$ in which almost all singularities
are not confining and such that $S^{(j)}({\bf{f}})\not\subset
S({\bf{f}}^{-k})$ for any positive $k$, then the
system is not algebraically integrable.}}\\
\\
As an example of this corollary, we show that System
(\ref{examplenc}), defined by
\begin{equation}
\label{examplencb} {\bf{f}}(x,y)= \left(
\begin{array}{c}
\displaystyle{-x-y+a+\frac{b}{x^3}}\\
x
\end{array}
\right),
\end{equation}
  is not algebraically integrable. Recall that  singularities of the form
$(0,y)$ are
  non confining. In order to use Corollary 5.1 and
prove that this mapping does not admit an algebraic first
integral, one must  show that points of the form $(0,y)$ are not
generically in the singular set of ${\bf{f}}^{-l}$ for some $l>0$.
The singular set of ${\bf{f}}^{-1}$ consists of points of the form
$(x,0)$. Thus, we must prove that generic iterates of $(0,y)$
under $\f^{-1}$ do not belong to the set of points of the form
$(x,0)$. The inverse of the mapping
  (\ref{examplencb}) is given by
\begin{equation}
{\bf{f}}^{-1}(x,y)= \left(
\begin{array}{c}
y \\
\displaystyle{-x-y+a+\frac{b}{y^3}}
\end{array}
\right).
\end{equation}
This mapping can be obtained from ${\bf{f}}$ of
(\ref{examplencb}) by interchanging $x$ and $y$. Therefore, we can follow
the analysis performed in (\ref{nonconfinement}) to perform a Laurent
expansion in
$y$ to show that
\begin{equation}
{\bf{f}}^{-2}(x_0,y_0)=\left(
\begin{array}{c}b/y_0^3+(a-x_0)-y_0+
\mathcal{O}({y_0^5})\\ -b/y_0^3+x_0+ \mathcal{O}({y_0^5})
\end{array}
\right).
\end{equation}
  Moreover, we have
\begin{equation}
{\bf{f}}^{-3}\left(
\begin{array}{l}
nb/y_0^3+(a-x_0)- y_0+\mathcal{O}({y_0^5})
\\
\\
    -nb/y_0^3+x_0+
\mathcal{O}({y_0^5})
\end{array}
\right)
= \left(
\begin{array}{l}
(n+1)b/y_0^3+(a-x_0)- y_0+\mathcal{O}({y_0^5})
\\
\\
-(n+1)b/y_0^3+x_0+ \mathcal{O}({y_0^5})
\end{array}
\right),
\end{equation}
where $n$ is any positive integer.
Hence, points of the form $(0,y)$ are not
generically sent
to points of the form $(x,0)$ by applications
of ${\bf{f}}^{-1}$
and, from Corollary 5.1, System (\ref{examplenc})
does not admit an
algebraic first integral.

In general,  algebraic integrability restricts the possible
\textit{local} behavior of  singularities.  Theorem 4.3
gives  two possible local behaviors but in each cases,
the singularity can have several different \textit{global} behaviors.
Consider an  irreducible component of the singular set
$S^{(1)}({\bf{f}})$. In the proofs of Theorems 3.3 and 4.3, we considered
two  cases: $S^{(1)}({\bf{f}})\not\subset \Lcv$ and
$S^{(1)}({\bf{f}})\subset \Lcv$. When
$S^{(1)}({\bf{f}})\not\subset \Lcv$ the singularities of
$S_I^{(1)}({\bf{f}})$ are generically confined. In this case, two
different types of global behavior can be expected. First, after
  meeting a finite number singularities, the mapping generically
never meets other singularities (note that this must also be true
in the backward direction, that is for ${\bf{f}}^{-1}$). We will
referred to  such a behavior as {\it{global confinement}}. Second, the
singularities may be  confined but not globally. Because, for each
dynamical system of the form (\ref{mapping}), there are only finitely
many irreducible parts to $S({\bf{f}})$, the mapping must, at one
point, come back to the initial singularity and take
$S_I^{(1)}({\bf{f}})$ into itself. We say that such singularities are
{\it{periodic}}. We now define formally these two types of
global behavior and illustrate them.\\
\\
{\noindent}{\bf Definition.} {\it{A singularity $\xb^* \in
\mathbb{C}^p$ for a
birational mapping $\xbu={\bf{f}}{(\xbn)}$ is
said to be {\bf{globally confined}} if there exist $k',k'' \in
\mathbb{N}^{>0}$ such that $\xb^* \not\in S({\bf{f}}^{k})$ for any
$k>k'$ and
$\xb^* \not\in S({\bf{f}}^{-k})$ for any $k>k''$.}}\\
\\
An example of such a behavior is given by System (\ref{3.0.1}).
For this system,  points of the form $(x,0)$ are not sent to
points of the form $(0,y)$ through successive applications of
${\bf{f}}$. To prove this, let $(x_n,y_n)$ be the $n$th iterate of
$(x,0)$. If $n=4k$, an expansion of   $\mathbf{f}^4(x,0)$ reveals
that $(x_{4k},y_{4k})=( k x+ \mathcal{O}(x^2), (-k+1) x+
\mathcal{O}(x^2))$, $k=1,2,\ldots$. This fact also implies that
$x_{4k+3}=-kx+ \mathcal{O}(x^2)$ and we conclude that $x_{4k+3}$
and $x_{4k}$ are neither infinite nor vanishing. By contradiction
assume that there exists $n$ such that $(x_n,y_n)=(0,y)$.
Therefore, either $n=4k+1$ or $n=4k+2$. However, from
Equation~(\ref{x22}) we know that $x_{n+2}$ is infinite which is a
contradiction. We conclude that no iterate of  $(x,0)$ falls onto
a point of the form $(0,y)$
and  singularities of the form $(0,y)$ are globally confined.\\
\\
{\noindent}{\bf Definition.} {\it{Consider the singular set  a
birational mapping $\xbu={\bf{f}}{(\xbn)}$ and one of its
irreducible component $S^{(j)}({\bf{f}})$. If there is a $k\in
\mathbb{N}^{>0}$ such that
${\bf{f}}^k(S^{(j)}({\bf{f}}))=S^{(j)}({\bf{f}})$, then the
elements of $S^{(j)}({\bf{f}})$ are said to be {\bf{periodic
singularities}} of period $k$.}}\\
\\
An example of such a behavior is given by a particular case of the
Gambier system \cite{11,12}
\begin{equation}
\left(
\begin{array}{c}
x_{n+1}\\
y_{n+1}
\end{array}
\right) = {\bf{f}}(x_n, y_{n}) = \left(
\begin{array}{c}
\displaystyle{\frac{(y_n+(1/a-a-1))x_n-a+1}{x_n+y_n}}\\
\\
\displaystyle{1-1/y_n}
\end{array}
\right), \label{periodic}
\end{equation}
where $a$ is a nonzero complex constant. The system
(\ref{periodic}) admits a singularity of type II at any point of
the form $(x,a)$. Then the first iterate of a point of this form
is given by $(1/a-1,1-1/a)$. But the Jacobians of ${\bf{f}}^2$ and
${\bf{f}}^3$ at $(x,a)$ are nonzero and the singularity is
confined. It is not globally confined since
${\bf{f}}^4(x,a)={\bf{f}}(x,a)=(1/a-1,1-1/a)$ and the singularity
is periodic. The system admit the following first integral
\begin{equation}
I(x,y)=\frac{y^3-3 y+1}{y(y-1)}.
\end{equation}

Now, consider the second case considered in  the proofs of
Theorems 3.3 and 4.3: $S^{(1)}({\bf{f}})\subset \Lcv$ for some
${\bf{c}}$. If the singularities of $S^{(1)}({\bf{f}})$ are
generically confined in $k$ steps, then
${\bf{f}}^k(S^{(1)}({\bf{f}}))$ has to be a set of the form of the
RHS of (\ref{inverse}). Since there are only finitely many of
those, the singularities cannot be globally confined. So, either
they are periodic, or $S^{(1)}({\bf{f}})$ lies inside the singular
set of ${\bf{f}}^k$ for all $k$ greater than a certain $k'$.  It
must also be true in the backward direction and if the
singularities are not periodic $S^{(1)}({\bf{f}})$ lies inside the
singular set of ${\bf{f}}^{-k}$ for all $k$ greater than a certain
$k''$. We will refer to these singularities as \textit{ubiquitous
singularities}. An example of such a behavior is given by the
singularities of Equation (\ref{exetrange}) of the  form $(x,0)$.
These singularities  are not confined but they  are in the
singular set of $\f^{-l}$ for any $l>0$ and therefore, these
singularities are ubiquitous.  Note that singularities of the form
$(0,y)$ are also ubiquitous. \\
\\
{\noindent}{\bf Definition.} {\it{A singularity $\xb^* \in
\mathbb{C}^p$  of a  birational mapping is
said to be {\bf{ubiquitous}} if there exist a $k' \in
\mathbb{N}^{>0}$ such that $\xb^* \in S({\bf{f}}^{k})$ for any
$k>k'$ and a $k'' \in \mathbb{N}^{>0}$ such that
$\xb^* \in S({\bf{f}}^{-k})$ for any $k>k''$.}}\\

Given a discrete dynamical system of the form (\ref{mapping}) and
an irreducible component $S^{(j)}({\bf{f}})$ of the singular set,
one would like to be able to obtain information on the first
integrals. The only way one can know if a given irreducible
component of the singular set lies inside a level set  is by
studying the global behavior of the singularity. If the
singularity is globally confined, then $S^{(j)}({\bf{f}})$ does
not lie inside a level set, but if it is ubiquitous,
then it does. If the singularities are periodic, one cannot
reach a conclusion.

Information on first integrals can be obtained by studying
singularities of type I. Consider the paths ${\bf{p}}_i$ defined
in the 2-dimensional case in the proof of Theorem 3.1. The number
of these paths is bounded above by the maximum of the degrees of
$P$ and $Q$. Moreover, for each $S_I^{(j)}({\bf{f}})\not\subset
\Lc$ for all $c$, consider $k'_j$, the lowest value $k_j$ can
take. Then $k'_j-1$ is the number of different paths ${\bf{p}}_i$
realized by iterating a point close to the singularity under the
mapping.
Hence, the following corollary holds.\\
\\
{\noindent}{\bf{Corollary 5.2.}} {\it{Consider a 2-dimensional
birational mapping with rational first integral $R=P/Q$.
Consider also the irreducible components of the singular set of the
first type
$S_I^{(j)}({\bf{f}})$ for which the singularities are globally
confined. Then the maximum of the degrees of $P$ and $Q$ is bounded
below by the number
\begin{equation}
\sum_j (k'_j-1),
\end{equation}
where $k'_j$ is given by the lowest value $k_j$ of Theorem 3.1.}}\\
\\
In Example (\ref{3.0.1}), the singular set has a unique
component in which singularities are globally confined with
$k=4$ and the degree of the first integral is exactly $k-1=3$. Note,
however, that the corollary only provides a  lower bound as shown in
the following example
\begin{equation}
\left(
\begin{array}{c}
x_{n+1}\\
y_{n+1}
\end{array}
\right) = \f(x_n, y_{n}) = \left(
\begin{array}{c}
\displaystyle{-y_n+x_n+\frac{a}{x_n}}\\
\\
\displaystyle{x_n}
\end{array}
\right), \label{lowerbound}
\end{equation}
where $a$ is a non-vanishing complex number. Here again, one can check
that singularities of the form $(0,y)$ are confined in 4 steps and
since   confinement is global, the lower bound for the
numerator and denominator of a first integral is 3. However, the
rational first integral $I=P/Q$ with lowest degree possible on $P$ and
$Q$ is given by
\begin{equation}
I=(x_n-y_n)^2(x_ny_n-a)^2,
\end{equation}
which is of degree 6, not 3 (note however the rather degenerate nature
of the first integral).

Next, we turn our attention to singularities of type II in the case
when
$S_{II}^{(j)}({\bf{f}})\not\subset \Lcv$ for all ${\bf{c}}$. From
the proof of Theorems 3.2 and 4.2, one sees that, as long as the
singularities are not confined, the  iterates of
$S_{II}^{(j)}({\bf{f}})\not\subset \Lcv$ must be roots of both
$P_i$ and $Q_i$ for all rational first integrals $P_i/Q_i$.
Therefore, we  have the following corollary.\\
\\
{\noindent}{\bf{Corollary 5.3.}} {\it{
Consider an algebraically
integrable $p$-dimensional birational mapping with the rational
first integral $R=P/Q$. Consider also an irreducible component of the
singular set of the second type $S_{II}^{(j)}({\bf{f}})$ whose
singularities are globally confined. Let $k_j$ be the corresponding
confinement number. Then the finite elements of the set
$$
\bigcup_{i=1}^{k_j-1} {\bf{f}}^i(S_I^{(j)}({\bf{f}})).
$$
are roots of both $P$ and $Q$.}}\\
\\
This corollary is best illustrated on  Example (\ref{dexample}) where
$(0,a)$ and $(1/a,0)$ are roots of both the numerator and
denominator of the rational function (\ref{integralfordexample}).

The previous corollary has an immediate consequence for the
existence of polynomial first integrals.\\
\\
{\noindent}{\bf{Corollary 5.4.}} {\it{Consider an algebraically
integrable $p$-dimensional birational mapping. If this system admits
singularities of the second type which are globally confined, then
none of the first integrals are polynomial.}}\\
\\
The last result concerns the case  when the irreducible component
of the singular set lies inside the closure of a level set of the
first integral. In Sections 3 and 4, it was found that if
${\bf{f}}^k$ (for $k\in \mathbb{N}$) is well-defined on
$S^{(j)}({\bf{f}})$ and its Jacobian is nonzero, then
${\bf{f}}^k(S^{(j)}({\bf{f}}))$ must be a set of the form of the
RHS of (\ref{inverse}), hence it lies on a level set. This is
stated in a following corollary.\\
\\
{\noindent}{\bf{Corollary 5.5.}} {\it{Consider a $p$-dimensional
birational mapping  $\xbu={\bf{f}}{(\xbn)}$ together with an
irreducible component of the singular set $S^{(j)}({\bf{f}})$ for which
the singularities are ubiquitous. If
${\bf{f}}^k$ is well-defined almost everywhere on
$S^{(j)}({\bf{f}})$ and its Jacobian is nonzero, then
${\bf{f}}^k(S^{(j)}({\bf{f}}))$ lies inside the closure of a level
set of the first integrals. Note that the case $k=0$ should also
be included meaning that $S^{(j)}({\bf{f}})$ itself lies inside
the closure of a level set of the first integrals.}}\\
\\
This corollary can be applied to Example (\ref{exetrange}). Since
the two sets of singularities ($(x,0)$ and $(0,y)$) are ubiquitous ,
they are level sets of the first integral
(\ref{firstintegraletrange}).

As a final example, consider the system
\begin{equation}
\left(
\begin{array}{c}
x_{n+1}\\
y_{n+1}
\end{array}
\right) = \f(x_n, y_{n}) = \left(
\begin{array}{c}
\displaystyle{-\yn+\xn+\frac{a}{\xn^2}}\\
\displaystyle{\xn}
\end{array}
\right), \label{counterexample}
\end{equation}
where $a$ is a nonzero complex number. This example was first used
in \cite{13}  to show that singularity confinement is not
sufficient for integrability. Following Example (\ref{3.0.1}), it
is easy to show that  singularities of the form $(0,y)$ are
generically globally confined. However, numerical analysis
performed on this system strongly suggest that is exhibits
chaotic behavior \cite{13}.


\section{Conclusions}\label{s6}

In this paper, we defined different types of
singularities and their confinement property for autonomous
discrete birational dynamical systems. The theory
of singularity confinement was developed more than 10 years ago
\cite{7,7a,7b} and different types of singularities have been
implicitly used. Here, in order to relate algebraic integrability to
singularity confinement we have found necessary to  give a
formal definition of different singularities and their confinement
property.

  In Sections III and IV, we showed that algebraic integrability for
birational mappings imply that singularities are either locally
confined or some preimages are not well-defined.

In Section V, we studied global properties of birational mappings and
showed that there are three types of behavior for singularities
compatible with algebraic integrability. Namely, singularities are
either globally confined, periodic, or ubiquitous. Remarkably,
  in the seminal papers describing singularity confinement, similar
behaviors were found heuristically to be compatible with integrability
\cite{7,7a,7b,9,11,12a}. However, the classical approach relies mostly
on a local analysis of the singularities. An important aspect of
the  results shown in this paper is that global properties have
to be satisfied for a system to be algebraically integrable.  To unify
these different concepts, we define the confinement property for a
discrete system in the following way:\\
\\
   {\noindent}{\bf Definition.} {\it{A birational mapping is said to
have the {\bf{confinement property}} if
   its singularities are, generically,  globally confined,
   periodic, or ubiquitous.}}\\
   \\
   Then the most important result of this paper can be rephrased  as
follows.\\
\\
{\noindent}{\bf Theorem.} {\it{An algebraically integrable  birational
mapping  has the confinement property.}}\\
\\
In particular, this result shows that every mapping in the
QRT-family \cite{QRT89,12a} has the confinement property both in
the symmetric and asymmetric cases. This fact also holds for the
recently discovered class of integrable birational discrete
systems discovered in \cite{KiYaHiRaGrOh02}. Moreover, from
examples such as System (\ref{counterexample}), we have strong
numerical evidence that the confinement property is necessary but
not sufficient for algebraic integrability.

In this paper, we also found sufficient conditions for
non-integrability. Essentially,  if singularities are confined in one
direction and not in the other, the system cannot be algebraically
integrable.

Finally, the remaining corollaries of Section 5 provide
information on the existence and degree of rational first
integrals directly from the singularity confinement procedure.

All the examples of integrable discrete systems considered in this
article have a first integral of genus 0 or 1. It is not possible
to find a birational mapping of infinite order preserving an
algebraic curve of genus 2 or higher. The argument proving this
fact is due to Veselov (\cite{Ve91}, page 35). It makes use of
Hurwitz theorem which tells us that the automorphic group of an
algebraic curve of genus 2 or higher is finite.

Since the only mappings studied here were birational, we could
have made use of the projective space in which the point at
infinity plays no particular role. Hence only singularities of
type II should be studied. However, the ideas presented in this
paper could not be extended for the case of non-autonomous or for
discrete dynamical systems that are not birational. For instance,
the integrable non rational mapping derived in
\cite{IaRo01,IaRo02,IaRo03} could not be studied. It is therefore
crucial to give a definition of singularity confinement that can
be applied to general discrete systems.

 Another important question concerns the relation between the confinement
property and other types of integrability for discrete dynamical
systems (see for example \cite{AbHaHe00,14,8,13,RoJoVi03}).


\begin{center}
{\bf{Acknowledgment}}
\end{center}
A.G. acknowledges a Fellowship from the Sloan Foundation. S.L.
acknowledges a post-doctoral fellowship from NSERC (National
Science and Engineering Research Council of Canada). The authors
are grateful to Basil Grammaticos, Alfred Ramani and John Roberts
for helpful discussions.

\end{document}